\documentclass[letterpaper,10pt]{article}

\usepackage{amssymb}
\usepackage{amsmath}
\usepackage{psfig}

\def\superspacyformat{
\setlength{\textheight}{8in}
\setlength{\textwidth}{5.7in}
\setlength{\evensidemargin}{0.2in}
\setlength{\oddsidemargin}{0.2in}
\setlength{\headheight}{0in}
\setlength{\headsep}{20pt}
\setlength{\topsep}{0in}
\setlength{\topmargin}{0.0in}
\setlength{\itemsep}{10pt}
\setlength{\parindent}{0pt}
\parskip=0.080in
}

\superspacyformat

\begin{document}

%A
\title{..........................................................ncp-0504-2002 \\
%A
.\\
%A
Theory of gravitation in flat space with no infinities}

%B\title{..................................................................................ncp-0504-20002\\
%B.\\
%BA Model of Elementary Particle Interactions}

\author{Irshadullah Khan \\
\\
National Center for Physics, Quaid Azam University, Islamabad\\
\\
irshadk@earthlink.net}
%A
\maketitle

%A
.\\
%A
\\
%A
\\
%A
\\
%A
\\
%A
\\
%A
\\
%A
\\
%A
\\
%A
\\
%A
\\
%A
\\
%A
\\
%A
\\
%A
\\
%A
\\
%A
\\
%A
\\
\newpage

%B\maketitle
%-------------------------------------------------------------------------
%-------------------------------------------------------------------------
\section{Introduction}

Nearly three decades ago an explanation for the cosmological redshift different from that usually accepted was 
presented \cite{Khan2}, \cite{Khan1}. It has not received much attention so far. Meanwhile, the standard explanation
for the redshift based on solutions of Einstein's equations \cite{Einstein1} applied to the cosmos has been the generally 
accepted  view on the subject. These solutions, derived by Friedmann  \cite{Friedman1}, \cite{Friedman2}  seem to suggest that the redshift is 
due to a universal cosmic expansion. Direct evidence against this viewpoint, however, was provided by the dissident 
astronomer Halton Arp \cite{Arp}, \cite{Bertola}.
% He was quietly ignored and was refused permission to continue his
% observational work which contributed to his retirement \cite{Hoyle}.
This evidence provided by Arp \cite{Arp}, \cite{Bertola} consists of 
observed close physical association between various cosmic objects with vastly different redshifts. The argument for close physical
association between the objects is nothing extra-ordinary either. It is quite simply the presence of visible filamentary 
bridges linking the objects. These objects
% called GRINDERS (GRross Non-DEpendance of distance on Red Shift) \cite{Khan3}, 
pose a basic challenge to standard cosmology.

No compelling evidence has been found for the link between curvature and gravity \cite{Einstein1} despite claims to the
contrary. In support of this viewpoint I would like to draw the reader's attention to the following facts.

1) The three standard consequences of the link between curvature and gravity are usually believed to be

a) observation of the anomalous perihelion advance of the planet Mercury, b) observed bending of star-light passing
close to the sun at a time of solar eclipse, c) apparent delay in transit time of radio pulses reflected from
planets when these pulses pass near the sun.

These three results are, however, derivable from a theory with flat space in which the speed of light and the rest 
mass at any point of space are functions of the Newtonian gravitational potential at that point. The various physical 
quantities remain finite everywhere even
when the Newtonian potential blows up at the location of a point mass. This theory is therefore as complete and simple as  
the standard theory claiming a link between gravitation and space time curvature. Moreover this theory has the attractive 
feature that no infinities 
in physical quantities occur at the location of point masses. Thus there are no black holes and all the paradoxes associated with this
enigmatic concept encountered elsewhere \cite{Thorne} \cite{Hawking} are now absent. I do not regard this as a shortcoming 
of the theory since no direct evidence has so far 
been presented for the occurrence of black holes. 
The evidence for black holes that has been presented  \cite{web2} involves
(a) the assumption of a limiting value for stellar masses (the so called Chandrasekhar limits for stars composed of
various known kinds of matter) necessitated by the singular nature of gravity in Newtonian and Einsteinian theories.
(b) application to 
observations of binary stars the Doppler shift formula derived under the assumption of local special relativity (LSR). The latter assumption 
becomes untenable according to 
the viewpoint presented in a separate communication \cite{Khan4}. The LSR Doppler shift  formula is to be replaced by a different formula
derived in \cite{Khan4}. Since the LSR formula has also been used in analyzing observations of the Taylor Hulse pulsar 
\cite{Taylor1} \cite{Taylor2} the claim of precise agreement between the actual motion of the binary star associated with 
the pulsar and the equations of general relativity becomes disputable. This theory of gravitation in flat space will 
presently be described in the following.

2) Because of the arguments presented in  (1), it is natural to turn to cosmic scales for any possible
evidence of the above mentioned link between curvature and gravity embodied in Einstein's theory of gravity \cite{Einstein1}. Again,
the evidence against the connection is overwhelming. Up to the farthest reaches of the universe available to our observing 
instruments, including the Hubble telescope,
the universe is absolutely flat \cite{Riess}, \cite{Garnavich}, \cite{Perlmutter}. This has resulted in all the efforts, 
seeking to avoid large scale curvature, that have led to modifications of Einstein's equations 
with terms arising from fields producing so called inflationary effects \cite{Guth}, \cite{Linde}, \cite{Steinhardt}
and the so called cosmological constant term \cite{Einstein3}.
None of these endeavors have received any justification in terms of what we know about fundamental forces besides gravity
\cite{Weinberg1}.

\section{Postulates and their consequences}

The flat space theory of gravitation starts from the following postulates:

The Lagrangian $(L)$ describing the motion of a body of rest mass (at infinity) equal to $m_0$ moving in the gravitational 
field of another body with rest mass at infinity $M_0>>m_0$, which is not acted upon by any force besides gravity, is
mathematically analogous to that of a freely moving body of mass $m_0$ in special relativity . It is given by
\begin{eqnarray}
\label{eqn1}
L & =  & -m c^2 \sqrt{1-{v}^{2}/c^2}
\end{eqnarray}

\begin{eqnarray}
\label{eqn2}
\mbox{where\,\,\,\,\,\,\,} c &  =  &c_0\; exp\;(-\frac{a G M_0}{c_0^2 r})
\end{eqnarray}
\begin{eqnarray}
\label{eqn3}
m &  = & m_0\; exp\;(\frac{b G M_0}{c_0^2 r})
\end{eqnarray}

$r$ is the distance between locations of the point objects whose rest masses at infinite separation are $m_0$ and $M_0$.
$c_0$ is the speed of light at an infinite distance from $M_0$. The presence of object with rest mass at infinity $= m_0$ 
is assumed to make a negligible contribution to variations in $c$ in the following derivation. $G $ is the Newtonian 
constant of gravitation = $6.67*10^{-10} Nm^2/kg^2$.

Introduce polar co-ordinates ($r, \theta$) in the plane of motion with the polar axis ($\theta = 0$) passing through the
origin at $M_0$. The principle of stationary action \cite{Goldstein} associated with the Lagrangian (\ref{eqn1}) leads to 
the equations of motion and gives rise to the constant of motion $H$ given by 

\begin{eqnarray}
\label{eqn4}
m r^2 \frac{d\theta}{dt} \sqrt{1+ \frac{{m_0}^2 {H}^2(u^2 +(u')^{2})}{ m^2 c^2}}& = & H m_0
\end{eqnarray}

where $u = \frac{1}{r},\,\, u' = \frac{du}{d\theta}$ and $u$ satisfies the following equation

\begin{eqnarray}
\label{eqn5}
u'' + u & = & \frac{G M_0}{H^2}\, \frac{m^2 c^2}{m_0^2 c_0^2}\, (2a-b) + \frac{GM_0}{c_0^2} \,a \,(u^2 +(u')^2)
\end{eqnarray} 

In order that this equation may reduce to the Newtonian limit when $ r\rightarrow\infty $, we must have

\begin{eqnarray}
\label{eqn6}
2a - b& =&1
\end{eqnarray}

So far equations (\ref{eqn4} ), ( \ref{eqn5}) are exact.

\subsection{Perihelion advance of planets}

We now develop a calculation involving a perturbation expansion around the approximate solution $u = u_0$ 
accurate up to the 0th order in $\frac{GM_0}{c_0^2}\,\, (\frac{GM_0}{H^2})$. \\
$u_0$ satisfies the equation

\begin{eqnarray}
\label{eqn7}
u_0'' + u_0 & = & \frac{1}{L}\\
L  & = &  \frac{H^2} {G M_0}
\end{eqnarray}

The solution $u_0 = \frac{1}{L}( 1+ e\, \cos(\theta))$ of equation (\ref{eqn7}) upon substitution in the terms of 
equation (\ref{eqn5}) that are of the 1st order in $\frac{GM_0} {c_0^2 L}$
gives the equation for the 1st order correction ($\frac{G M_0}{c_0^2 L}\, u_1$) to the approximate solution 
$u_0$. The equation for $u_1$ is thus

\begin{eqnarray}
\label{eqn8}
u_1'' + u_1 & = & \frac{1}{L}\,[a(1+e^2) - 2(a-b)(2a-b )] + \frac{e}{L}\, \cos(\theta)\,[2a-2(a-b)(2a-b)]
\end{eqnarray}

From equation (\ref{eqn8}) it now follows that the perihelion advance per revolution is approximately

\begin{eqnarray}
\label{eqn9}
&\delta \theta = 2\pi \,\frac{G\,M_0} {c_0^2 L} b&
\end{eqnarray}
\begin{eqnarray}
&L = {\cal A} \,\,(1-e^2)&
\end{eqnarray}

where $\cal{A}$ is the semi major axis of the orbit.

In order that for the planet Mercury, after allowing for perturbations by other planets, the residual perihelion 
advance of 43" per century, corresponding to astronomical observations (Leverrier), may become explicable 
by this first order correction to the Newtonian theory one must have

\begin{eqnarray}
\label{eqn91}
b& = &3+ \epsilon  
\end{eqnarray}

where $\epsilon$ is a small number. From equations (\ref{eqn6} ),  (\ref{eqn91})  it then follows that

\begin{eqnarray}
\label{eqn92}
a& = & 2+ \frac{1}{2} \epsilon
\end{eqnarray}

Notice that with the values for $a$ and $b$ given by (\ref{eqn92}) and (\ref{eqn91}), eqns. (\ref{eqn2}), (\ref{eqn3})
imply that 

\begin{eqnarray}
&m(r) \,c^2(r) \,\rightarrow \, 0 \,\, \mbox{as} \,\,r \,\rightarrow \,0&
\end{eqnarray}

Thus if $\,m(r)\,c^2(r)\,$ is interpreted as the total energy content of the test mass $m(r)$ when it is stationary at a location
that is at distance $r$ from $M_0$ then this interpretation would lead to the conclusion that when an
active external agent is acting upon the test mass causing it to move quasi-statically from $r=\infty $ to $r=0$ the
total energy gained by the agent from the test mass is $m_0\,c_0^2$. This is precisely the entire energy content of
the test mass at $\infty$ according to the theory of special relativity. Hence when the test mass eventually disappears
in the manner just described it does so without taking away any energy that becomes inaccessible because of what is called 
an event horizon in the theory of general relativity \cite{Wald}.

\subsection{Bending of star light by the sun}

The time taken by light to travel along a path connecting two fixed points $A$ and $B$ is 

\begin{eqnarray}
\label{eqn93}
\frac{1}{c_0}\,{\int_A}^B n(r)\, \,ds & = & \frac{1}{c_0}\,{\int_A}^B \,n(r)\,\sqrt {(\frac{dr}{d\theta})^2 + r^2 } \,\,d\theta
\end{eqnarray}

where  $n(r)= c_0/c(r)$ is the refractive index of light at a distance $r$ from the sun

According to Fermat's principle of stationary time the path followed by light in going from $A$ to $B$ must be such that
the time taken is stationary under variations of the path. This leads to the following differential equation for the path of a light ray.

\begin{eqnarray}
\label{eqn10}
n(r)\, \frac{d^2r}{d{\theta}^2}- n'(r)\, \left( r^2 + (\frac{dr}{d\theta})^2 \right) - \frac{n(r)}{r}\, \left( r^2 + 2 
(\frac{dr}{d\theta})^2\right)=0
\end{eqnarray}

where the $'$ now denotes differentiation with respect to $r$.
Substituting

\begin{eqnarray}
\label{eqn10a}
z= (1/r^2)\, (\frac{dr}{d\theta})^2 +1
\end{eqnarray}

in (\ref{eqn10} ) one gets

\begin{eqnarray}
\label{eqn10b}
z'/z & = & 2(n'/n  +1/r) 
\end{eqnarray}

Equation (\ref{eqn10b}) integrates to

\begin{eqnarray}
\label{eqn10c}
z & = & (1/K^2)\, r^2 \,n(r)^2
\end{eqnarray}

so that equations (\ref{eqn10a}) and (\ref{eqn10c}) give

\begin{eqnarray}
\label{eqn10d}
\frac{dr}{d\theta} = -(1/K )\,r \,\sqrt{n(r)^2 r^2- K^2}
\end{eqnarray}

The change in the direction of a ray of starlight incoming from $r = \infty$ to the point $P$ where $\frac{dr}{d\theta}
=0 $, $\, \,i.e. \,\, r=p\, \, \mbox{ such that } n(p)\, \,p = K $,  is 

\begin{eqnarray}
 \phi & = &- K {\int_\infty}^p \frac{1} {r \sqrt{(n(r)^2 r^2 - n(p)^2 p^2)}}\,\,\,\, dr   - \frac{\pi}{2}
\end{eqnarray}

The change in direction from $P$ to infinity along the other branch of the path is equal to the above expression
so that the total change noticed by terrestrial observers is nearly

\begin{eqnarray}
\label{eqn11}
2\,  \phi &=& (2+ \frac{1}{2}\,\epsilon)(\pi-1)\,\frac{G M_0}{c_0^2 p}
\end{eqnarray}

If $\epsilon = 0$ then this result (\ref{eqn11}) for the deflection of star light would be greater by about 7 per cent than that calculated according
to Einstein's theory \cite{Weber} according to which 

\begin{eqnarray}
\label{eqn11a}
2\, {\phi}_{G.R.} &=& \frac{4GM_0}{c_0^2 p}
\end{eqnarray}

Weber \cite{Weber} has mentioned that the average for eleven eclipses is 
in agreement with (\ref{eqn11a} ) to roughly 0.2 per cent. However I believe that this difference does not
rule out the present theory because the influence of the plasma (ionized matter) surrounding the sun on light
propagation near the sun must also be considered. The presence of a plasma whose density decreases with
increasing distance from the sun (as expected) causes the effective refractive index to become less than its value in the 
absence of the plasma. Since this effect is frequency dependent it could in principle be separated from the effect due
to gravity.

\subsection{Time delay for pulsed radiation grazing the sun (Shapiro experiment)}

The time delay for an electromagnetic pulse passing near the boundary of the solar disk (solar limb) can now 
be calculated. It is given by

\begin{eqnarray}
\label{eqn12}
\mbox{time delay}& = &2( {\int_p}^{D_1}\frac{ds}{dr}\,\frac{dr}{c(r)} +{\int_p}^{D_2}\frac{ds}{dr} \frac{dr}{c(r)} - (\sqrt{{D_1}^2-p^2} 
+\sqrt{{D_2}^2-p^2})/c_0)
\end{eqnarray}

where $D_1$ and $ D_2 $ are the distances of the planet (reflecting the pulse) and the earth from the sun and
$p$ is the shortest distance from the sun to the path of the radio pulse. 

From equation (\ref{eqn10d}) it follows that

\begin{eqnarray}
\label{eqn12a}
\frac{ ds}{dr}= \sqrt{1+ r^2 (\frac{d\theta}{dr})^2} = \frac{n(r) r}{\sqrt{n(r)^2 r^2 - K^2}}
\end{eqnarray}

Substituting for $\frac{ds}{dr}$ from (\ref{eqn12a}) into (\ref{eqn12}) and
retaining terms upto the first order in $G \,M_0/c_0^3$ equation (\ref{eqn12}) may be approximated as 

\begin{eqnarray*}
\mbox{time delay}=
 \frac{2}{c_0} &&\left({\int_p}^{D_1} +{\int_p}^{D_2} dr\, (1 -\frac{p^2}{r^2})^{-\frac{1}{2}}\,(1 +\frac{2 G\,M_0}{c_0^2 r} -
\frac{2 G\,M_0}{c_0^2} \frac{p}{r(r+p)})\right. 
\end{eqnarray*}
\begin{eqnarray}
\label{eqn12b}
&\left.-\sqrt{D_1^2 -p^2} - \sqrt{D_2^2-p^2}\right)&
\end{eqnarray}

This expression for the time delay (\ref{eqn12b}) may be further simplified using the relation $p << D_1, D_2$ to give

\begin{eqnarray}
\label{eqn13}
\mbox{time delay}& = & 4 G\,M_0\,(-2 + ln(D_1 D_2/p^2))(1 + \frac{\epsilon}{4})
\end{eqnarray}

This result (\ref{eqn13}) may be compared with the following expression (\ref{eqn14}) derived from the equations of general relativity (G.R.) with 
the specific choice of co-ordinates giving the Schwarzschild form  of the space-time metric \cite{Weinberg3}
{\em and the assumption that the distances of planets from the sun are given by the Schwarzschild radial co-ordinate
despite its singular behavior near the sun}. This expression (\ref{eqn14}) forms the basis
of Shapiro's analysis \cite{Shapiro} for time delay of radar echoes from the planets.

\begin{eqnarray}
\label{eqn14}
\mbox{time delay}_{G.R.} & = & 4 G\,M_0\, (1 + ln(D_1 D_2/p^2))
\end{eqnarray}

If $\epsilon=0$ then the difference between (\ref{eqn13}) and (\ref{eqn14}) applied to the case of Mercury  
is close to the uncertainty in the arrival time of the pulse. This uncertainty is due to the spreading of the pulse.
Because of this "the experiment presents extraordinary difficulties in execution and interpretation" \cite{Weinberg3}.  
Both the results (\ref{eqn13}) (with $\epsilon =0$) and (\ref{eqn14}) are in agreement with observations at the present time.

\section{Implications of the Pound-Rebka and the Hafele-Keating  experiments}

I next turn to the physical meaning of the Pound-Rebka experiment \cite{Pound} in the context of the present viewpoint regarding
variation of $c$ and $m$ with $r$ given by equations (\ref{eqn2}) and (\ref{eqn3}). Let us recall a simplified version of 
what is being observed in the Pound-Rebka experiment. In this experiment recoil-less emission of gamma radiation by nuclei 
(Mossbauer \cite{Mossbauer1}) is observed. The emission occurs with negligible recoil of the nucleus because the nucleus is 
firmly embedded in the surrounding crystalline lattice of atoms of which it is an integral part. This has the consequence 
that the emitted gamma has a very narrow width with no Doppler broadening expected of nuclei in, for example, a gas of 
molecules in thermal agitation.

Because of the sharpness of the gamma lines these lines may be used as a probe of very small variations in their 
characteristics through various kinds of changes in the circumstances under which these lines may be generated and/or
detected. Thus the Pound-Rebka experiment consists of investigating resonant absorption of these lines produced by a source
when they pass through an absorber (which is a much smaller version of the source itself) where the source and 
absorber are arranged to be located at different heights in the gravitational field of the earth. The observation made by
Pound and Rebka may be described essentially as follows. {\em In order that the absorber may produce maximal absorption of
the gammas it must be moving towards (away) from the source with a speed proportional to its height (depth) above (below) the source with the 
constant of proportionality equal to $g/c$ where g is the acceleration due to gravity and c is the speed of light, both 
measured on the surface of earth.}

To analyze the entire situation very carefully, leaving no room for error, let us imagine two identical stationary atoms one
at ground level (location 1) at a distance $R$ from the center of earth and the other at location 2 which is at a height 
$\delta R$ ($<< R$) above the ground level. Let us denote the energy content of two atomic states $a, b$ to be $E_{a1},
E_{b1}: (E_{b1} >E_{a1})$ respectively for the stationary atom in location 1 and $E_{a2},E_{b2}$ respectively for the 
similar atom in location 2. The frequency of the photon produced in the transition $b \rightarrow a $ occurring at 
location 1 is $\nu_1(b_1;a_1)$ so that the photon energy content at location 1 ($E_1(b_1;a_1$)) is given by

\begin{eqnarray}
\label{eqn14a}
E_1(b_1;a_1)=h_1 \,\nu_1(b_1;a_1) = E_{b1}-E_{a1}
\end{eqnarray}

where $h_1$ is the value of Planck's constant at location 1.

The wave length ($\lambda_1(b_1; a_1)$) of the electromagnetic wave associated with the photons produced in the transition 
$b \rightarrow a$ at location 1 is then 

\begin{eqnarray}
\label{eqn14b}
\lambda_1(b_1;a_1)= c_1/\nu_1(b_1;a_1)
\end{eqnarray}

where $c_1$ is the velocity of light at location 1 given by eqn(\ref{eqn2}). Analogous equations , obtained via the substitution
of 1 by 2 in (\ref{eqn14a}, \ref{eqn14b}), hold for the photons produced in the transition $b \rightarrow a$ at
location 2. Actual experiments involve the transport of
photons (or devices governed by photons) from one location to another and comparison with the photons generated 
(or devices governed by the local photons) at the other location. Let us therefore call the frequency and wavelength 
of the photon 1 (generated in the transition $b \rightarrow a$ of an atom at location 1) after its arrival at location 2 
to be $\nu_2(b_1;a_1)$ and $\lambda_2(b_1;a_1)$ respectively. After its arrival at location 2 its behavior is {\em assumed to 
be indistinguishable from the locally generated photons}. 
Therefore

\begin{eqnarray}
\label{eqn14c}
\lambda_2(b_1;a_1)= c_2/\nu_2(b_1;a_1)
\end{eqnarray}

\begin{eqnarray}
\label{eqn14d}
E_2(b_1;a_1)=h_2 \nu_2(b_1;a_1)
\end{eqnarray}

where $E_2 (b_1; a_1)$ is the energy of the photon (generated at location 1) after its arrival at location 2, $c_2$ is the velocity of light and $h_2$ is the Planck constant at location 2.

Now it may be shown from the hypothesis of mass energy equivalence that when a stationary atom at location 1 in state $a$ or $b$ is lifted quasi-statically against gravity to
location 2, with no kinetic energy imparted to it, the energy content is increased by the factor $(1 + g\, \delta R/c_0^2)$.
Thus we have

\begin{eqnarray}
\label{eqn15}
E_{a2}/E_{a1} = & E_{b2}/E_{b1} & = 1 + g\, \delta R/c_0^2
\end{eqnarray}

Equations (\ref{eqn2}, \ref{eqn92} ) give

\begin{eqnarray}
\label{eqn16}
c_2/c_1 & = & 1 + (2 +\frac{\epsilon}{2}) g \,\delta R/c_0^2
\end{eqnarray}

Observation of spectral lines from the sun imply that their wavelengths (when compared with those of similar lines generated 
by stationary atoms on earth ) are increased on arrival at earth. The magnitude of this increase in wave length for
lines in the solar spectrum observed from earth suggests that if a similar experiment were performed on earth, 
i.e. the experiment in which photons generated in the transition $b \rightarrow a$ at location 1 were allowed to 
propagate to location 2 and then compared with the photons generated in the transition $b \rightarrow a$ at location 2,
one would find

\begin{eqnarray}
\label{eqn17}
\frac{\lambda_2(b_1;a_1)}{\lambda_2(b_2;a_2)}& = & 1 + g \,\delta R/c_0^2
\end{eqnarray}

Equations (\ref{eqn14c}, \ref{eqn17}) combined with the equation obtained from (\ref{eqn14b} ) through substitution of
1 by 2 then give

\begin{eqnarray}
\label{eqn18}
\frac{\nu_2(b_1;a_1)}{\nu_2(b_2;a_2)} = \frac{\lambda_2(b_2;a_2)}{\lambda_2(b_1;a_1)} = 1- g \,\delta R/c_0^2
\end{eqnarray}

The Pound Rebka experiment \cite{Pound}, \cite{Cranshaw} has quite independently demonstrated that the photon arriving from the lower level (location 1) to the upper
level (location 2) needs to be boosted in energy and frequency 
by the factor $(1+g \delta R/c_0^2)$ in order that its energy may be changed from $E_{b1}-E_{a1}$ to $E_{b2} - E_{a2}$ and it may be resonantly 
absorbed by atoms at location 2. Hence

\begin{eqnarray}
\label{eqn19}
\nu_2(b_2;a_2) & = & (1 + g \,\delta R/c_0^2)  \nu_2(b_1;a_1)
\end{eqnarray}

This relation (\ref{eqn19} ) is thus seen to be in agreement with equation ( \ref{eqn18}) suggested by terrestrial observation of spectral
lines from the sun as well as equation (\ref{eqn15}) derived from the mass energy equivalence.

We shall now turn to the only other experiment that has a bearing on the issues being discussed here. The problematic 
aspects of this experiment due to Hafele and Keating \cite{HK} will be presented in the next section. 
In the Hafele Keating experiment \cite{HK} a cesium clock flown in a commercial airliner was compared with another
at ground level. It was claimed that for the airborne clock the number of its ticks had received a positive contribution 
due to its height ($\delta R$). The amount of the claimed contribution indicates that

\begin{eqnarray}
\label{eqn20}
\frac{\nu_2(b_2;a_2)}{\nu_1(b_1;a_1)} & = & 1+ g \,\delta R/c_0^2
\end{eqnarray}

\subsection{ Critical remarks about the Hafele Keating experiment and their consequences}

The above analysis of Hafele Keating experiment ( \ref{eqn20} ) would have been unambiguously sound if the following considerations
were not involved .  Hafele and Keating have also claimed in their paper \cite{HK} the existence of a contribution to the proper time difference between the airborne 
clock and another at  ground level which is dependent on the velocity of the aircraft and which vanishes as this velocity 
goes to zero. In other words, according to their analysis, the circum-navigation with very slow velocities (quasi-static
motion) does not produce any mis-match between the two clocks at the end of the journey. This assertion is certainly
in contradiction with the form of the metric in a rotating system given by a succession of investigators such as Einstein
\cite{Einstein1}, Landau \cite{Landau} and also with the somewhat different form of the metric suggested by Post\cite{Post}.
According to these authors global synchronization of clocks on rotating frames is impossible and  in such frames circum-navigation
of a clock moving quasi-statically along a closed path is expected to produced a mis-match in time ($\delta T$) with the
stationary clock given by

\begin{eqnarray}
\label{eqn20a}
\delta T &=& \pm \frac{2\,S \,\omega}{c^2}
\end{eqnarray} 

where $\omega$ is the angular velocity of rotation and $S$ is the area enclosed by the path of the circum-navigating clock (or the path of counter-propagating light 
beams constrained by  mirrors in the Sagnac experiment \cite{Sagnac} ).
Notice that this mis-match between a circum-navigating and a stationary clock depends on the velocity of rotation of 
earth and is absent from the Hafele Keating \cite{HK} analysis.
However what makes this result (\ref{eqn20a}) important for us to retain is that it also provides a simple understanding 
of the Sagnac effect \cite{Sagnac}. In the Sagnac experiment two counter-propagating electromagnetic waves along a closed 
path defined by a fixed set of
mirrors in a rotating frame emerge with a phase difference. The Sagnac effect has been observed both in the case 
(a) when the source as well as observer looking at the fringes are in the rotating frame \cite{Sagnac}  and also in the case (b) when the source
and the observer are in a (near inertial) frame with respect to which the set of mirrors defining the closed path is 
rotating \cite{Dufour1}, \cite{Dufour2}, \cite{Marinov}.  In case (b) the effect is understood as arising from the 
difference in lengths of the two paths traversed by light traveling
with the same speed along them. In both cases the frequency of the light moving along the two paths is assumed to be 
the same. In case (a) the lengths of the paths are the same and the observed phase difference is due to the time difference
at the end of circum-navigation given by  $2\, \delta T$ (\ref{eqn20a}). Eqn. (\ref{eqn20a}) has been derived from the 
transformation to rotating frames of the
Lorentzian  space-time metric \cite{Landau} \cite{Einstein1}.

If this term (\ref{eqn20a}) is really absent as Hafele and Keating seem to suggest from  analysis of their experiment then in the
first version of the Sagnac effect experiment (which is the version directly relevant to the Hafele Keating experiment)
we would have to say that the phase difference is due to a difference in the frequency of light along the two paths. 
Moreover we would also have to accept that this difference in frequencies of light along the two paths is not manifested 
by an analogous behavior involving clocks since Hafele and Keating do not seem to need it for their analysis of their
experiment \cite{HK}. So photons would not be analogous to clocks in their time keeping behavior in rotating frames.

Another aspect of their analysis for the time difference that needs to be carefully addressed is the following. They relate
the proper times for the two clocks to a hypothetical clock located in an imagined near-inertial frame at rest relative to the center
of the earth. After ascertaining the difference in proper times from that perspective they convert it back to
the clock
located at ground level using local special relativistic formulas since the difference in proper times of the two clocks
is measured on the ground level clock in the actual experiment. So the natural question to be asked is what happens
to their calculation if the same local special relativistic formulas are used but the proper times recorded by the two
clocks are related to a hypothetical clock located at the center of the earth in an imaginary near-inertial frame {\em moving 
uniformly} relative to that in which the center of the earth is at rest. An analogous analysis then shows that the result of the 
calculation is changed in the general case of an arbitrary inertial frame in uniform motion with respect to the special frame chosen 
in the Hafele Keating analysis. So if their analysis is to be accepted as correct, one must accept Hafele and Keating's 
implicit viewpoint that not all inertial frames are equivalent \cite{HK}

In view of the above remarks about their analysis it is plausible to suggest that perhaps the error estimates quoted by Hafele and Keating are
much smaller than the errors actually incurred. If this is accepted as a correct statement then the Sagnac term (\ref{eqn20a})
and the gravitational red shift term together may completely account for their data. 

However, we shall take the precautionary viewpoint that the Hafele Keating experiment is inconclusive as  suggested by 

a) contradiction with the usually accepted explanation of the Sagnac effect \cite{Sagnac}, \cite{Dufour1}, \cite{Dufour2}- see
above. This effect has been observed time and again by different workers and is the basis of technological
devices for keeping track of fixed directions in aircraft and Hitachi cars. 

b) adopting a different explanation for the Sagnac effect (as described above) from the
usually accepted one would require us to accept time keeping behavior of clocks quite different from those of photons.

c) an explanation for the entire effect in the Hafele Keating experiment \cite{HK} along the lines suggested by Hafele and
Keating \cite{HK} requires that inertial frames are not all equivalent. 

d) no repetition of the experiment has been undertaken by these authors or any one else during the past thirty years.

e) In the present framework of \,ideas, it is in disagreement with Einstein's principle of equivalence for light 
propagation in accelerated frames (to be explained below: see equation (\ref{eqn22b}))

We shall therefore replace equation (\ref{eqn20} ) by

\begin{eqnarray}
\label{eqn21}
\frac{\nu_2(b_2;a_2)}{\nu_1(b_1;a_1)} & = & 1+ A\, g \,\delta R/c_0^2
\end{eqnarray}

The constant $A$ would then have to wait for its un-ambiguous determination in a future better version of the Hafele Keating 
experiment or another experiment described below.

\subsection{Proposed experiments}

Now I propose that the total energy content of the photon generated at location 1 is the same after its arrival
at location 2. This is because, unlike the atom which has a non zero rest mass, non-gravitational forces do not intervene 
to bring a photon to rest at one location from rest at another location through the application of non-gravitational forces.
A photon is never at rest. So if non-gravitational forces have not made their appearance during the propulsion of the 
photon under the influence of the conservative force of gravity then its energy content ($h \nu$) must stay constant just as the
total energy content of an atom moving under the influence of gravity alone would stay constant. There is a viewpoint 
\cite{Weinberg2} 
in the literature contrary to this. This viewpoint is based on the argument of making a distinction between the kinetic
energy and potential energy of the photon and
identifying the former only with $h\nu$. The question of the appearance and disappearance of the potential energy of 
the photon during its emission from and absorption by the atom is thus completely ignored. Clearly then this viewpoint is
tantamount to the suggestion that the total  effective 
energy content of the photon does not stay constant during its motion in the conservation field of gravity. This 
argument would then seem to suggest that the photon is not subject to the conservative nature of gravity unlike the other
particles. We propose to avoid this viewpoint and its conclusion.   

Therefore for a photon generated at location 1 and arriving at location 2 equality of energy content at the two locations
implies

\begin{eqnarray}
\label{eqn22c}
h_1 \,\nu_1(b_1;a_1) & = & h_2 \,\nu_2(b_1;a_1)
\end{eqnarray}

Equation (\ref{eqn22c}) combined with the result of the Pound-Rebka experiment  (\ref{eqn19} ) and equation (\ref{eqn21} )
then implies

\begin{eqnarray}
\label{eqn22}
h_1/h_2 = \frac{\nu_2(b_1;a_1)}{\nu_1(b_1;a_1)}=\frac{\nu_2(b_1;a_1)}{\nu_2(b_2;a_2)}\, \frac{\nu_2(b_2;a_2)}{\nu_1(b_1;a_1)}
=1+ (A-1)g\, \delta R/c_0^2
\end{eqnarray}
\begin{eqnarray}
\label{eqn22a}
\mbox{i.e. \,\,\,\,\,\,}\nu_1(b_1;a_1) & = & (1- (A-1) g \,\delta R/c_0^2)\,\nu_2(b_1;a_1)
\end{eqnarray}

Two comments are in order here.

(a) If the conclusion (\ref{eqn20}) of the Hafele Keating analysis \cite{HK} is sound in spite of the problematic aspects
of the other assumption in their paper (as explained above) then $A =1$ and $h_1 =h_2$. 

(b) If \,Einstein's principle of equivalence, applied to light propagation in
accelerated frames holds then eqn. (\ref{eqn22a}) implies that

\begin{eqnarray}
\label{eqn22b}
 A=0.  &\mbox{\,and\,}  & h_1 \neq h_2
\end{eqnarray}

Now according to Bohr's quantum theory the atomic energy content is proportional to $\alpha^2 m c^2$ where $\alpha$ is the
fine structure constant and $m$ is the mass of the electron. 

Hence eqn. (\ref{eqn15}) gives,

\begin{eqnarray}
\label{eqn23}
\frac {\alpha_1^2 m_1 c_1^2 }{\alpha_2^2 m_2 c_2^2}& = E_{a1} / E_{a2} = &1 -g\, \delta R/c_0^2
\end{eqnarray}

Equation (\ref{eqn3}) implies

\begin{eqnarray}
\label{eqn24} 
m_1/m_2 & = & 1 + (3 +\epsilon)\, g \,\delta R/c_0^2
\end{eqnarray}

Equations (\ref{eqn16}), (\ref{eqn23}) and (\ref{eqn24}) lead to the conclusion

\begin{eqnarray}
\label{eqn25}
\alpha_1 & = & \alpha_2
\end{eqnarray}

Equations (\ref{eqn16}), (\ref{eqn22}), (\ref{eqn25}) now  imply that

\begin{eqnarray}
e_2/e_1 = 1 + \frac{1}{2}\,(3+\frac{\epsilon}{2}-A)g \,\delta R /c_0^2 
\end{eqnarray}

Thus one  expects to find  an electric field in the regions above and below two parallel metal plates perpendicular to the 
direction of gravity and charged to a potential difference $V_2 - V_1$. For the ideal case of infinite plates the magnitude of 
this electric field is calculated to be

\begin{eqnarray}
&\frac{|(3+\epsilon/{2}-A)| \, g \, |V_2-V_1|}{2\, c_0^2}&
\end{eqnarray}

The direction of the electric field in the region above the plates is along (opposite to) the direction of gravity
according as $(3+\epsilon/2-A)(V_2-V_1)\, \mbox{is} \,< 0 \,(> 0)$.  In the region below the plates the direction of the electric field
is opposite. Since according to our present approach to physical theory the possibility of an apparently fundamental distinction
between uniform gravitational field and a uniformly accelerated frame has been hinted at by the inconclusive Hafele Keating 
experiment (equations \ref{eqn20}, \ref{eqn21}, \ref{eqn22b}) it is conceivable that this distinction might become manifest 
in certain other experiments.

The principle of equivalence may now be directly tested by the following experiment. Take two coaxial metal cylinders of 
radii $R_1$ and $R_2 > R_1$. Charge these cylinders by applying a potential difference $V_2 - V_1$  across the two cylinders.
Spin the cylinders about the common axis with the same constant angular velocity $\omega$.  Under the conditions of 
the experiment the charges on the two cylinders are of opposite sign and their quantity per unit length ($q$) is expected 
from application of equivalence principle to differ by 

\begin{eqnarray}
\label{eqnq}
&\delta q =(\frac{3}{4}+\frac{\epsilon}{8})\,q \omega^2 (R_2^2-R_1^2)/(c^2)&
\end{eqnarray}

Hence electric field $ E $ at a distance $R (> R_2)$ from the axis of symmetry is 

\begin{eqnarray}
\label{eqne}
E & = &(\frac{3}{4}+\frac{\epsilon}{8})\,\,\frac{V_2 -V_1}{ln(R_2/R_1)}\,\, \frac{\omega^2\, \,(R_2^2-R_1^2)}{c_0^2 \,R}
\end{eqnarray}

In (\ref{eqnq}) and (\ref{eqne})  we have assumed $A=0$ in order that the Einstein equivalence principle for light propagation in accelerated
frames \cite{Einstein1} may hold  (\ref{eqn22b}).
Finally let us mention that in order that charge may be attenuated by the force of gravity and the problems of
plus infinity and minus infinity associated with electric charges and gravitational masses respectively may both be 
overcome we must have $A-\frac{\epsilon}{2} < 3$. 

\section{Acknowledgements}

I would like to thank Professor Riazuddin, Director National Center for Physics, Islamabad for his kind hospitality.

I would also like to express my deep appreciation to Allan Widom, Fayyazuddin, Karen Fink, Yogi Srivastava, Mahjoub Taha, 
 Husseyin Yilmaz, Abdel-Malik AbderRahman, Mohammed Ahmed, Asghar Qadir, Qaisar Shafi, Abner Shimony,
Badri Aghassi, Peter Higgs, Derek Lawden, Robert Carey,  Priscilla Cushman, Stephen Reucroft, 
Tahira Nisar, John Swain, Alan Guth, Fritz Rohrlich, Martinus Veltman, Kalyan Mahanthappa, Claudio Rebbi, 
Stephen Maxwell, Sidney Coleman, Christian Fronsdal, Moshe Flato, John Strathdee, Henrik Bohr, Julian Chela-Flores,
Muneer Rashid, Klaus Buchner, Marita Krivda, Uhlrich Niederer, Werner Israel, Patricia Rorick, Daniel
Donkoh, Leroy Burrows, Yasushi Takahashi, Helmut Effinger, John Synge, John Wheeler, Nandor Balazs and the 
{\em departed souls: Abdus Salam, Marvin Friedman, Cornelius Lanczos, Asim Barut, Jill Mason, Sarwar Razmi, 
Nicholas Kemmer, Paul Dirac, Lochlainn O'Raifeartaigh, Peter and Ralph Lapwood, Iqbal Ahmad, Naseer Ahmad}
(may they be favored by Allah) for friendship and/or moral support and/or conversations on various occasions. 

I am grateful to my son Bilal for his cheerful involvement in my life.

\newpage
\bibliography{cosmos}

\end{document}